\documentclass[aps,prb,twocolumn,superscriptaddress,groupedaddress]{revtex4}
\usepackage{hyperref}
\usepackage{bm}
\usepackage{graphicx}
\usepackage{amsmath}
\usepackage{sidecap}
\usepackage{amsfonts}
\usepackage{amssymb}

\usepackage{color}

\begin{document}
	\title{Non-degenerate surface pair density wave in the Kagome superconductor CsV$_3$Sb$_5$ - application to vestigial orders}
	\author{Yue Yu}
	\affiliation{Department of Physics, Stanford University, Stanford, CA 94305}
	\affiliation{Department of Physics, University of Wisconsin, Milwaukee, WI 53201}
	\begin{abstract}
		On the Sb-layer of the Kagome superconductor CsV$_3$Sb$_5$, pair density wave states have been observed. When the high-temperature charge orderings are treated as static backgrounds, these PDW states exhibit the same wavevector in the effective 2D Brillouin zone. Interestingly, these PDW states break the same symmetry on the surface. Considering the presence of this non-degenerate PDW, we investigate the implications for the possible existence of a vestigial charge-4e phase with a non-zero center-of-mass momentum. To distinguish between different vestigial phases, we propose scanning tunneling microscopy experiments. We aim to provide insights into the nature of the vestigial phases and their distinct characteristics in CsV$_3$Sb$_5$. This research sheds light on the interplay between PDW states, charge orderings, and superconductivity of the Kagome superconductor.
	\end{abstract}
	
	\maketitle
	
	\section{ introduction}
	The recently discovered Kagome superconductor CsV$_3$Sb$_5$ (CVS) \cite{ortiz2019,Ortiz2020} is a highly intriguing material that has attracted significant experimental and theoretical interest due to its exotic charge and superconducting orderings \cite{jiang2023}.
	
	In its high-temperature phase, the CVS crystalizes in the P6/mmm space group and exhibits a layered structure composed of V-Sb sheets intercalated by Cs atoms \cite{ortiz2019}. At a temperature of 94K, scanning tunneling microscopy (STM) experiments have revealed a $2a_0$ charge density wave (CDW) state with a $2\times2$ superlattice modulation \cite{xu2021,chen2021,liang2021,wang2021}. Notably, the CDW has been found to possess a three-dimensional character, with reports of both $2\times2\times2$ and $2\times2\times4$ modulations \cite{Li2021,Miao2021,Ortiz2021,liang2021,song2022}. 
	
	Moreover, different intensity distributions of the CDW peaks with unusual magnetic response have been observed, indicating the presence of a chiral charge order \cite{jiang2021,zhao2021,wang2021}. Investigations are underway on the time-reversal symmetry breaking in this system \cite{mielke2022,yu2021,li2022No,feng2021Chiral,park2021,feng2021,denner2021,lin2021,yu2021Concurrence}. The difference in CDW intensity also suggests rotational symmetry breaking. The temperature dependence of the nematic transition has been recently explored \cite{xiang2021,ni2021,Li2021,ratcliff2021,wang2021Unconventional,luo2022possible}.

	Around 60K, an unidirectional CDW with a periodicity of $4a_0$ is observed on the Sb-layer through STM measurements \cite{zhao2021,liang2021,chen2021,li2023}. However, this CDW is not observed on the alkali layer or in the bulk \cite{Li2021,Miao2021,Ortiz2021,song2022}. Superconductivity is observed in CsV$_3$Sb$_5$ at a temperature of 2.3K \cite{Ortiz2020}, and the nature of its order parameter is still under investigation\cite{mu2021,yu2012,kiesel2013,wang2013,duan2021,luo2022,xu2021,zhao2021Nodal,wu2021,chen2021,gu2022,gupta2022,chen2021Double,yu2021unusual,chen2021highly,zhang2021}.
	
	In the context of the superconducting phase, the presence of pair density wave (PDW) on the Sb-layer has been reported by STM probes \cite{chen2021}. Similar to the $4a_0$ CDW, its existence on the alkali layer and in the bulk remains unconfirmed, resulting in unknown 3D wave vectors. Consequently, a comprehensive analysis of the bulk superconducting phase diagram becomes challenging. Conversely, the STM results offer valuable insights and provide sufficient information to construct the superconducting phase diagram on the surface. Therefore, studying the surface phase diagram can serve as a stepping stone towards a better understanding and analysis of the bulk superconducting phase.
	
	In this study, our focus is directed towards the surface superconducting transitions. To simplify the analysis, we consider the charge orderings that emerges at significantly higher temperatures than the superconductivity as a static background. This treatment effectively enlarges the unit cell and results in a folded Brillouin zone. Within this folded Brillouin zone, it is observed that all PDWs exhibit the same wavevector. These PDWs, being physically equivalent, can be effectively described by a single order parameter denoted as $\Delta_{\bf Q}$.
	
	The presence of the unusual PDW has significant implications for the vestigial phases. Traditionally, vestigial phases are constructed using two independent PDW order parameters\cite{berg2009}. However, in the case of the non-degenerate PDW $\Delta_{\bf Q}$ and the uniform SC $\Delta_{\bf 0}$ in CVS, both order parameters need to be considered to construct vestigial phases.
	
	Specifically, if a charge-4e phase exists, it would be characterized by the composite order parameter $\Delta^{4e}=\Delta_{\bf 0}\Delta_{\bf Q}$. The presence of such a phase can be examined through experimental techniques other than the Little-Parks oscillation\cite{ge2022}. In this study, we will utilize STM signatures to differentiate between different vestigial phases and provide insights into their distinct characteristics.
	
	This paper will primarily investigate the properties of the non-degenerate PDW state on the low-temperature Sb surface. A significant focus will be placed on presenting a symmetry argument that prohibits the existence of the conventional uniform charge-4e phase. Furthermore, a comparison of various vestigial phases will be conducted, emphasizing their distinct STM signatures. Additionally, the role of CDW disorder in stabilizing vestigial superconducting phases will be discussed, particularly in the context of commensurate systems. The necessity of CDW disorder for the stabilization of these vestigial phases will be explored and elucidated.

	\section{Symmetry Analysis}
\begin{figure}[h]
		\centering
		\includegraphics[width=5cm]{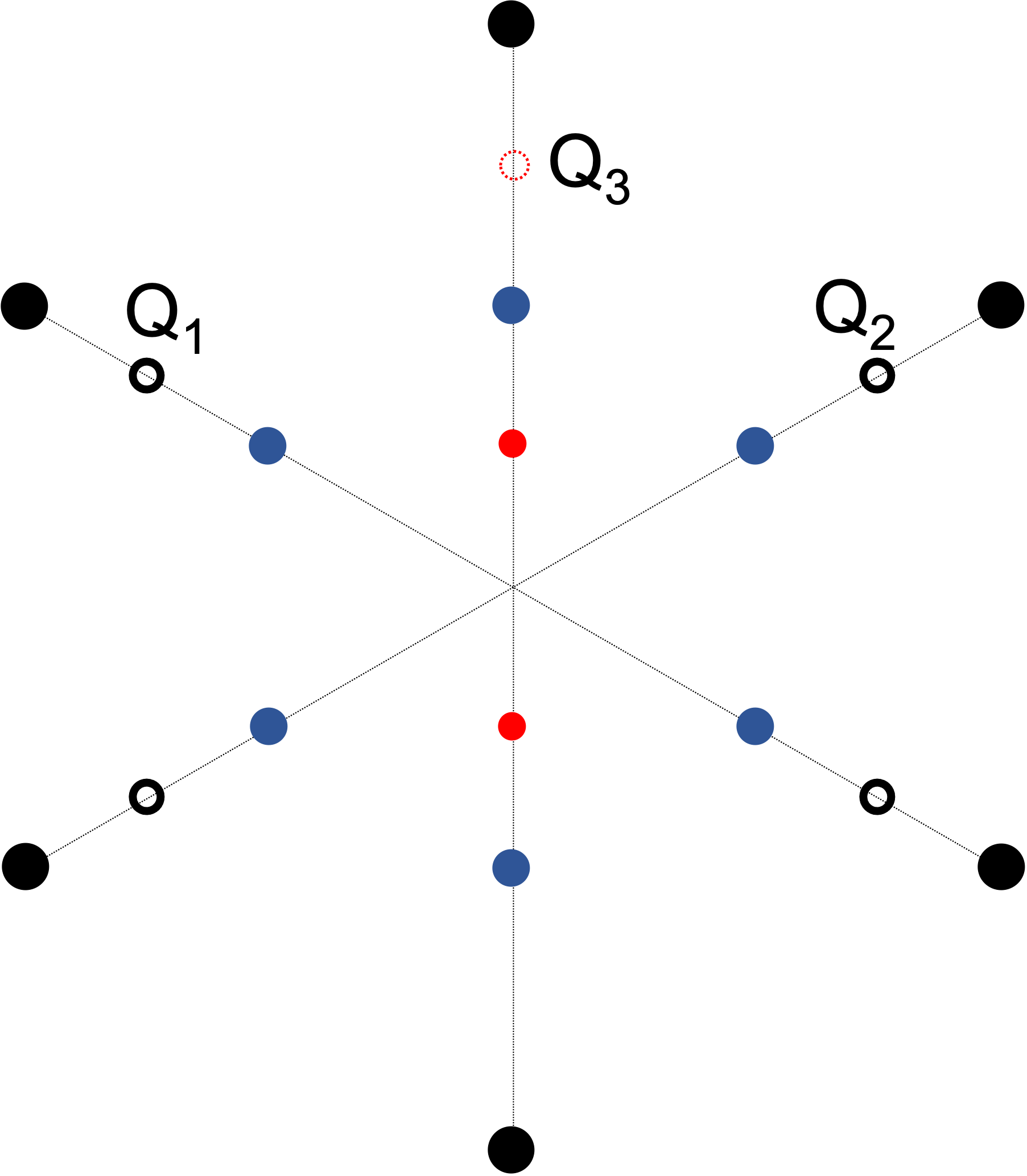}
		\caption{The Bragg peaks from the Kagome lattice structure (black), 
			$2a_0$-CDW (blue) and $4a_0$-CDW (red). Inside the superconducting phase, four additional peaks (black circles at $\bf\pm Q_{1,2}$) are observed in STM.}
		\label{F1}
	\end{figure}

	In this study, our focus is on the low-temperature PDW state observed on the Sb-cleaved surface of the Kagome superconductor CVS. On this surface, multiple CDW states have been reported above the superconducting phase. The first CDW state, with a $2a_0$ periodicity, is observed at 94K \cite{Ortiz2020,zhao2021,jiang2021, Li2021}. Its wavevectors, shown as blue dots in Fig.\ref{F1}, preserve the 6-fold rotational symmetry. The second CDW state, with a $4a_0$ periodicity in the XY-plane, emerges around 60K \cite{zhao2021,chen2021}. Its wavevectors, shown as red dots, break the 6-fold rotational symmetry. Both CDW ordering temperatures significantly exceed the superconducting transition temperature $T_c\approx2.3K$.
	
	Deep in the SC phase at 300mK, four additional CDW $\rho_{\bf Q_i}$ with peaks at $\bf\pm Q_{1,2}$ have been observed in STM \cite{chen2021}, which are the signature for PDW $\Delta_{\bf Q_i}$. Given the existence of a uniform SC component $\Delta_0$, these CDW  wave vectors $\bf Q_i$ are the same as the PDW wave vectors through $\rho_{\bf Q_i}=\Delta_0^*\Delta_{\bf Q_i}$. The same wave vectors are observed in the spatial variation of the SC gap magnitude\cite{chen2021}. The exact critical temperature for the low-temperature CDW is currently unknown. However, experimental evidence suggests that it is below 4.2K \cite{chen2021}. The peak at $\bf \pm Q_3$ is a higher harmonics of the unidirectional CDW, which exists even above the SC phase\cite{chen2021}, so it does not break additional translational symmetry. 
	
	\begin{figure}[h]
		\centering
		\includegraphics[width=6cm]{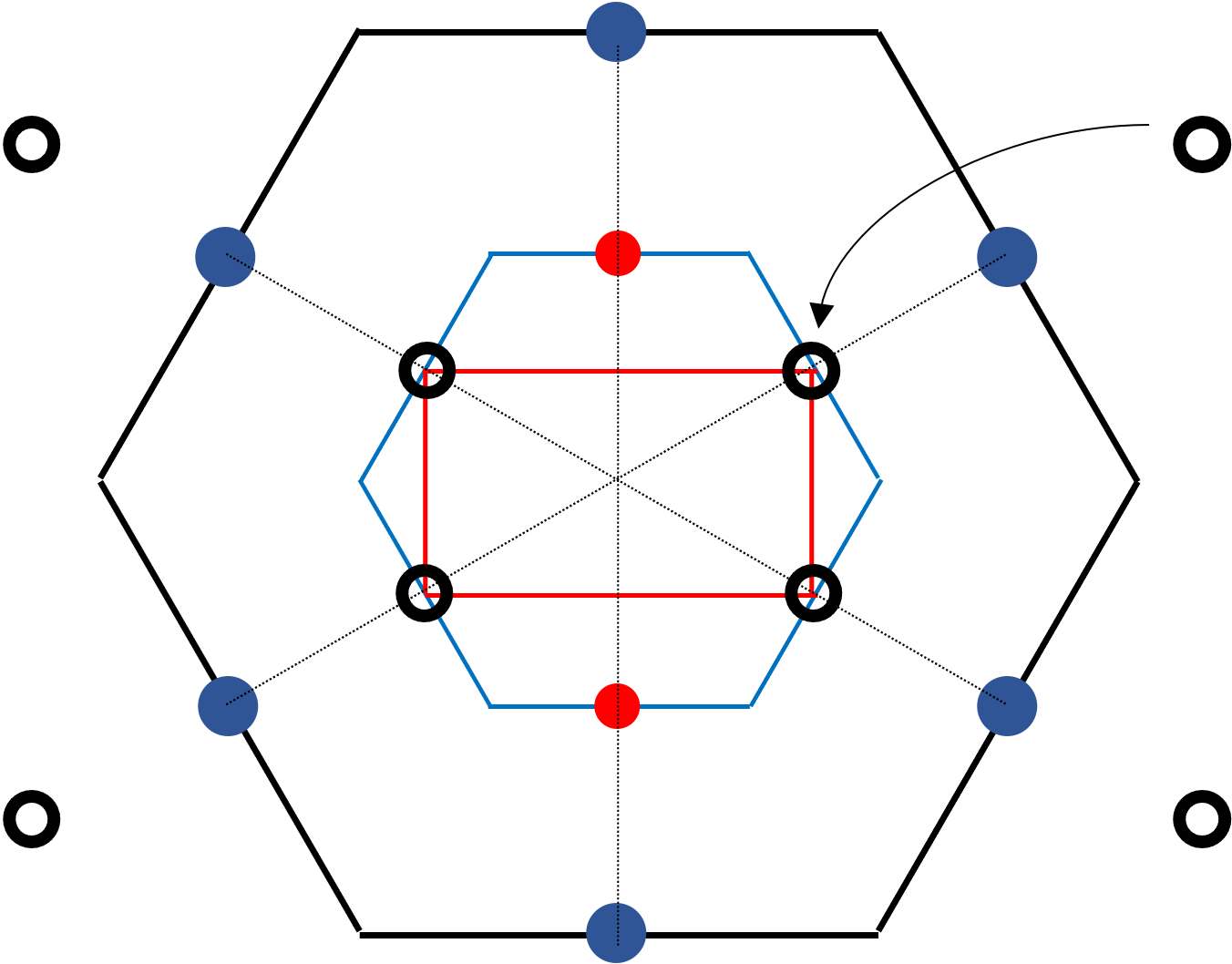}
		\caption{The unfolded Brillouin (black line), folded Brillouin zone due to $2a_0$-CDW (blue line) and folded Brillouin zone due to both $2a_0$ and $4a_0$-CDW (red line). In the last folded Brillouin zone, the PDW (black circle) is located at the corner. }
		\label{F2}
	\end{figure}
	
	When considering the low-temperature phases ($T<4K$) in the presence of the $2a_0$ and $4a_0$ CDW, it is appropriate to treat them as static backgrounds. This leads to a larger unit cell, requiring the Brillouin zone (BZ) folding. In Fig.\ref{F2}, the original Brillouin zone is depicted as a solid black hexagon. Upon folding the Brillouin zone according to the $2a_0$ CDW (blue dots), the resulting Brillouin zone (indicated by the blue line) remains a hexagon, but with a halved size. The PDWs (black circles) are now folded onto the M points of the resulting Brillouin zone. Consequently, the PDWs $\Delta_{\bf Q_i}$ and $\Delta_{\bf -Q_i}$ become physically identical, breaking the same symmetry.

    We now proceed to fold the BZ according to the unidirectional $4a_0$-CDW (red dots). The resulting folded Brillouin zone is depicted as a red rectangle. Interestingly, all four PDWs (originally $\bf \pm Q_{1,2}$) are now located at the corners of the rectangle (i.e. the $(\pi,\pi)$ point). This implies that these PDWs are physically equivalent, exhibiting the same symmetry breaking.
	
	To summarize, the low-temperature phase observed on the Sb-layer of CVS features a non-degenerate PDW $\Delta_{{\bf Q}=(\pi,\pi)}$ and a uniform superconducting order parameter $\Delta_{\bf 0}$. This differs from the well-known `LO'-like phase \cite{larkin1972}, which exhibits a doubly-degenerate PDW $\Delta_{\bf \pm Q}$. Furthermore, the low-temperature phase in CVS is distinct from the `FF'-like phase \cite{fulde1964} as the wavevector $\bf Q$ preserves time-reversal symmetry.

	\begin{figure}[h]
		\centering
		\includegraphics[width=8cm]{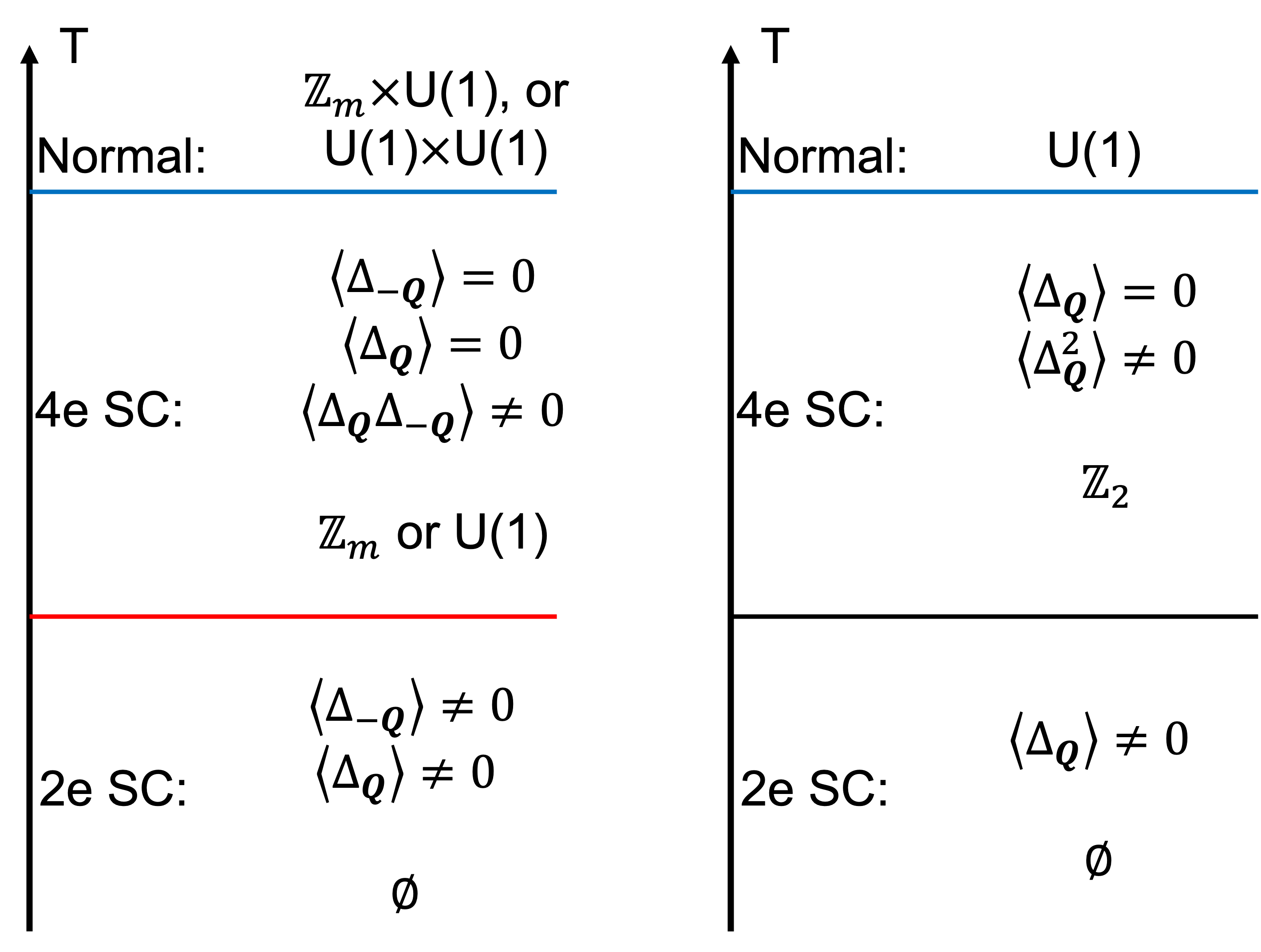}
		\caption{(Left) The conventional phase diagram hosting a vestigial charge-4e phase from doubly degenerate PDWs $\Delta_{\pm\bf Q}$\cite{berg2009}. The blue line denotes the superconducting transition while the red line denotes the CDW transition. The remaining symmetry in each phase is included. Depending on the commensurability of the CDW, CDW transition breaks U(1) (incommensurate) or $\mathbb{Z}_m$ (commensurate) symmetry. (Right) {\it Unlikely} phase diagram hosting a charge-4e phase using only a non-degenerate PDW. It cannot be achieved since the discrete symmetry breaking (black line) generically happens above the continuous symmetry breaking (blue line).}
		\label{F3} 
	\end{figure}
	
	The non-degenerate PDW $\Delta_{\bf Q}$ introduces an interesting scenario for vestigial phases. Traditionally, vestigial phases require two independent superconducting order parameters. In the case of the doubly degenerate PDW $\Delta_{\bf \pm Q}$, the vestigial charge-4e phase arises from the uniform order parameter $\Delta^{4e}_0=\Delta_{\bf Q}\Delta_{\bf -Q}$  \cite{berg2009}. In this phase, $\Delta^{4e}_0$ is ordered while the corresponding CDW $\rho_{2{\bf Q}}=\Delta_{\bf Q}\Delta_{\bf -Q}^*$ is disordered. The phase diagram is depicted in the left panel of Fig.\ref{F3}, where the blue line represents the superconducting transition and the red line represents the CDW transition. The remaining symmetries in each phase are also indicated.
    The superconducting transition breaks a U(1) symmetry, while the CDW transition can either break an additional U(1) symmetry if the CDW $\rho_{2{\bf Q}}$ is incommensurate or a $\mathbb{Z}_m$ symmetry if the periodicity of the CDW is m times the lattice spacing. To support the charge-4e phase, it is necessary for the superconducting transition to occur at a higher critical temperature than the CDW transition. In the case of an incommensurate CDW, this requirement can be met in a clean system as long as the superconducting phase has a larger stiffness \cite{berg2009}. However, for a commensurate CDW, discrete symmetry breaking generally occurs at a higher $T_c$ than continuous symmetry breaking in a clean system. In such cases, CDW disorder can be required to suppress the CDW $T_c$, allowing the charge-4e phase to persist.
	
	In the case of CVS, the conventional construction of the charge-4e phase is not applicable due to the presence of the $2a_0$-CDW, which makes $\Delta_{\bf \pm Q}$ the same order parameter. In principle, one could consider a charge-4e phase where $\Delta_{\bf Q}^2$ is ordered while $\Delta_{\bf Q}$ remains disordered. The corresponding phase diagram is depicted in the right panel of Fig.\ref{F3}. However, this phase cannot be stabilized. In this charge-4e phase, $\Delta_{\bf Q}$ becomes an Ising variable, leading to the breaking of the continuous U(1) symmetry into a $\mathbb{Z}_2$ symmetry. For this charge-4e phase to be realized, it would require the U(1) continuous symmetry breaking to occur at a higher temperature than the $\mathbb{Z}_2$ discrete symmetry breaking for the same order parameter. However, in practice, this is unlikely to happen since discrete symmetry breaking, which is less affected by fluctuations, typically occurs at higher temperatures than continuous symmetry breaking.

	Vestigial phases, if existed, should have alternative order parameters, constructed from two distinct SC orders. In this system, the choice is the PDW $\Delta_{\bf Q}$ and the uniform SC $\Delta_{0}$. The vestigial charge-4e order and the vestigial CDW order thus should be $\Delta^{4e}_{\bf Q}=\Delta_{0}\Delta_{\bf Q}$ and $\rho_{\bf Q}=\Delta_{0}\Delta^*_{\bf Q}$. 
	
\begin{figure}[h]
		\centering
		\includegraphics[width=8cm]{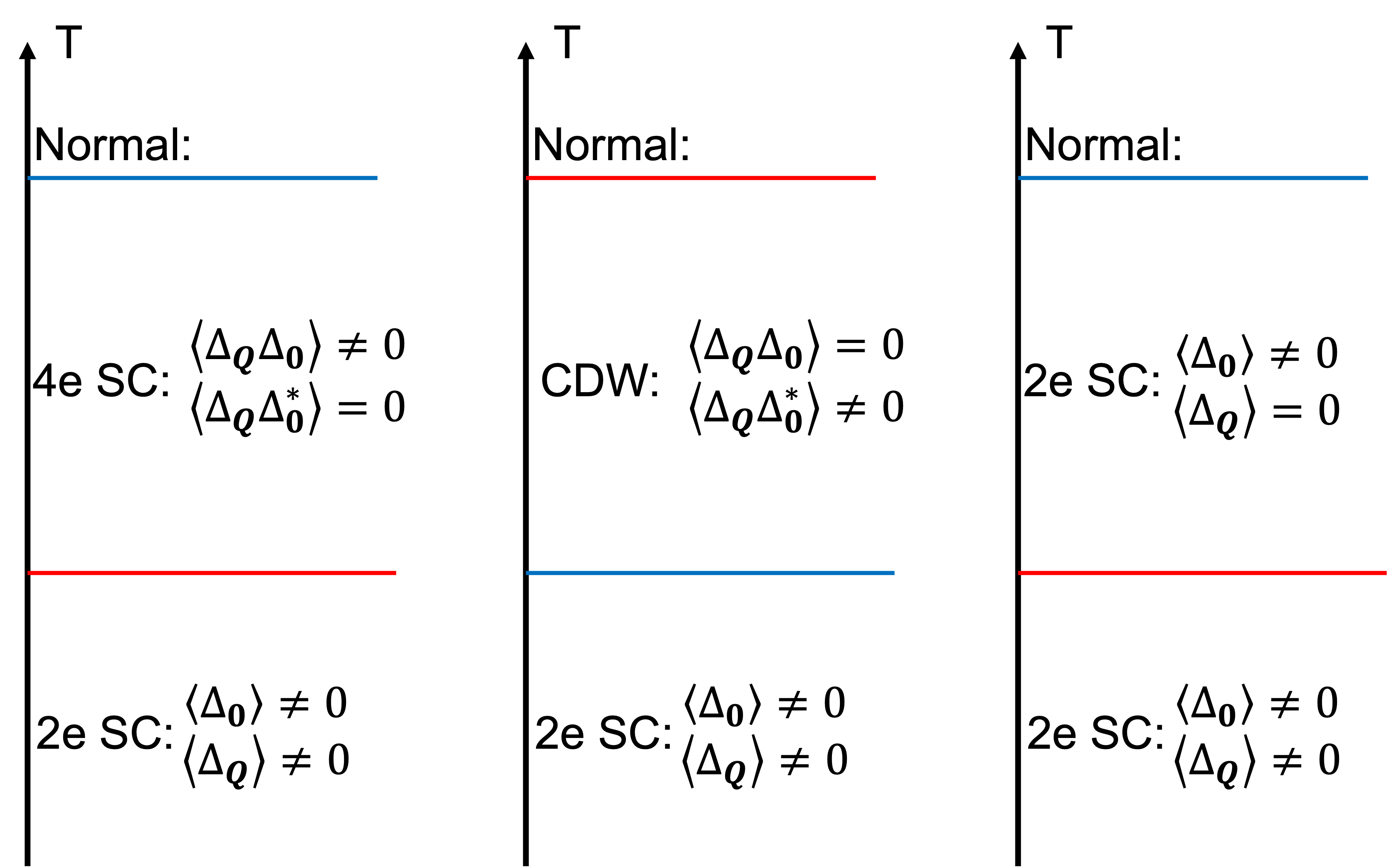}
		\caption{Possible phase diagrams with the uniform SC $\Delta_0$ and the non-degenerate PDW $\Delta_{\bf Q}$ at low temperature. The blue line is a U(1) superconducting transition while the red line is a $\mathbb{Z}_2$ CDW transition. (Left) The vestigial charge-4e SC phase is developed first, with $\Delta_{0}$ and $\Delta_{\bf Q}$ coexisting in the charge-2e SC phase. (Middle) The vestigial CDW phase is developed first, with $\Delta_{0}$ and $\Delta_{\bf Q}$ coexisting in the charge-2e SC phase. (Right) No vestigial phases are developed, with $\Delta_{0}$ and $\Delta_{\bf Q}$ ordered at different temperatures.}
		\label{F4} 
	\end{figure}

	As shown in the left panel of Fig.\ref{F4}, the phase diagram hosting the charge-4e phase has the same essence as the conventional phase diagram in Fig.\ref{F3}. The charge-4e SC transition (blue line) is a U(1) symmetry breaking, where $\Delta^{4e}_{\bf Q}=\Delta_{0}\Delta_{\bf Q}$ is ordered. The CDW transition (red line) is a $\mathbb{Z}_2$ (since ${2\bf Q}={\bf 0}$) translational symmetry breaking, where $\rho_{\bf Q}=\Delta_{\bf Q}\Delta_{0}^*$ is ordered. It is worth emphasizing that, the CDW here and below, corresponds to peaks at $\pm Q_{1,2}$. Its critical temperature is below 4.2K.
 To host the charge-4e phase, the continuous U(1) symmetry breaking needs to have a higher $T_c$ than the discrete $\mathbb{Z}_2$ symmetry breaking. This is difficult to achieve for a clean system. However, CDW disorder can suppress the CDW $T_c$, allowing the charge-4e phase to survive. A numerical analysis of the effect of the disorder can be found in Sec.\ref{S4}. 
	
	\section{Experimental signatures}
	From the symmetry argument in the previous section, vestigial phases need to have a composite order parameter from both $\Delta_{0}$ and $\Delta_{\bf Q}$. In this section, we will illustrate the experimental signatures of possible phase diagrams. 
	
	We will focus on the STM probe. In the charge-4e phase, there is a non-zero density of states for gapless excitations, due to the lack of $\bf Q=0$ pairing components. This has the essence as the previous studies on the doubly-degenerate PDW phase\cite{lee2014}. As usual, the STM gap opening, i.e. vanishing density of states at the Fermi level, is a signature for the ordering of $\Delta_0$. And the CDW peaks at $\pm\bf Q_{1,2}$ are signatures for the ordering of $\rho_{\bf Q}=\Delta_0\Delta_{\bf Q}^*$. We will use these two signatures to distinguish phase diagrams with/without vestigial phases.
	
	In the vestigial SC phase $\Delta^{4e}_{\bf Q}\neq0$ (left panel of Fig.\ref{F4}), the PDW and the uniform charge-2e order can mutually induce each other through the term $\Delta^{4e}_{\bf Q}\Delta^*_{0}\Delta^*_{\bf Q}$ in Landau theory. Consequently, the gap opening for $\Delta_{\bf 0}$ and $\Delta_{\bf Q}$ must happen at the same temperature. The STM gap opening and the CDW peaks at $\pm\bf Q_{1,2}$ will thus appear at the same temperature (red line).
	
	If the vestigial charge-4e phase does not exist, there could be a vestigial CDW phase (middle panel). In this phase, $\rho_{\bf Q}=\Delta_{\bf Q}\Delta_{\bf 0}^*$ is ordered, while SC is not. At the SC $T_c$, both gaps $\Delta_{\bf 0}$ and $\Delta_{\bf Q}$ open for the same reason above. In STM, the CDW peaks $\pm\bf Q_{1,2}$ (red line) will thus appear at a higher temperature than the SC gap opening (blue line). 
	
	If neither vestigial phases exist, then the gaps $\Delta_{\bf 0}$ and $\Delta_{\bf Q}$ will generically open at different temperatures, as shown in the right panel of Fig.\ref{F4}. We take $\Delta_0$ as a stronger ordering with a higher critical temperature as $|\Delta_0|$ is comparable to the superconducting $T_c$ \cite{chen2021}. In STM, SC gap opening (blue line) will appear at a higher temperature than CDW peaks $\pm\bf Q_{1,2}$ (red line).
	
	For simplicity, we only discussed the above three possibilities. Other possibilities (e.g. charge-6e phase) will lead to more complicated phase diagrams. As long as vestigial SC phases (4e or 6e) exist, $\Delta_0$ and $\Delta_{\bf Q}$ can induce each other in that phase. So the above properties (same $T_c$ for gap opening and CDW peaks) still provide a straightforward way to check the possibility of vestigial SC phases.
	
	Another more direct way to detect the charge-4e phase is by phase-sensitive measurement. In real space, the charge-4e order parameter reads: $\Delta^{4e}({\bf r})=|\Delta^{4e}_{\bf Q}|\exp(i\bf Q\cdot r)$. When putting this charge-4e SC next to a uniform charge-2e superconductor $\Delta^{2e}$, it will induce a PDW order $\Delta_{\bf Q}=\Delta^{4e}_{\bf Q}\Delta^{2e,*}$, as well as a CDW order $\rho_{\bf Q}=\Delta^{4e}_{\bf Q}\Delta^{2e,*}\Delta^{2e,*}$ at the boundary.

	
	\section{ Disorder Effect}\label{S4}
	For the vestigial SC state to appear, the critical temperature $T_{SC}$ needs to be higher than $T_{CDW}$. A commensurate CDW breaks discrete symmetry while SC breaks continuous symmetry. In a clean system, thermal fluctuation in SC would likely be stronger than CDW, leading to $T_{SC}<T_{CDW}$. In this case, CDW disorder can help suppress $T_{CDW}$, thus allowing the vestigial SC phase to survive. The effect can be described by the following Landau-Ginzburg-Wilson free energy density:
	\begin{equation}
		\begin{split}
			f({\bf r})&=k_{\Delta}|\nabla\Delta|^2+\frac{u_{\Delta}}{2}|\Delta|^2+\frac{1}{4}|\Delta|^4\\
			&+k_{\phi}(\nabla\phi)^2+\frac{u_{\phi}}{2}\phi^2+\frac{1}{4}\phi^4+\frac{\lambda}{2}|\Delta|^2\phi^2+h({\bf r})\phi
		\end{split}
	\end{equation}
	Here $\Delta$ is a complex field, describing the U(1) symmetry breaking for the charge-4e SC. $\phi$ is a real field, describing the $\mathbb{Z}_2$ symmetry breaking for the commensurate CDW near the superconducting transition. The $\lambda$ term describes the competition between them. 

 We include the CDW disorder as a real random field variable $h({\bf r})$, with the spatially independent $h({\bf r})$ uniformly distributed within $[-D, D]$. The parameter D then measures the disorder strength. Note that the SC cannot be coupled to any random-field disorder due to the gauge symmetry.
 
This minimal model captures the most straightforward consequence of CDW disorder, which is suppressing $T_{CDW}$ and thus allowing the vestigial SC phase. We find it unnecessary to include all symmetry-allowed terms, originated from the rotational symmetry breaking and the possible time-reversal symmetry breaking in the normal phase. For example, the inclusion of time-reversal symmetry breaking, represented as an internal magnetic field, can lead to the formation of magnetic vortices within the superconducting state.  The introduction of random-$T_c$ disorder on the superconductivity can also result in the presence of local SC patches, which can enhance the superconducting $T_c$ under proper conditions.
It is essential to clarify that our intention in this work is not to undermine the importance of these physics. Rather, our primary objective is to emphasize that, CDW disorder {\it can} help the existence of the vestigial SC phase through its most straightforward and non-exotic mechanism.

	The random-field disorder is known to destroy long-range ordering in 2D, but more realistically, we will focus on a finite-sized system.  We compute the correlation functions: $c_{\Delta}(r)=\frac{1}{L^2}\sum_{x,y}\langle\Delta(x,y)\Delta^*(x+r,y+r)\rangle$ and $c_{\phi}(r)=\frac{1}{L^2}\sum_{x,y}\langle\phi(x,y)\phi(x+r,y+r)\rangle$. Here $\langle\rangle$ denotes the thermal average. Then we obtain configurational average $\overline{c_{\Delta,\phi}(r)}$ over different disorder samplings. 
	
	For a $L\times L$ periodic system, we mainly look at the correlation functions at the longest distance $C_{\Delta,\phi}\equiv\overline{c_{\Delta,\phi}(L/2)}$. The phase boundaries can be determined when $C_{\Delta,\phi}$ reaches a threshold value. $u_\Delta=-1$, $u_\phi=-0.95$ are taken for the classical Monte Carlo calculation. Details can be found in the appendix.

	\begin{figure}[h]
		\centering
		\includegraphics[width=4cm]{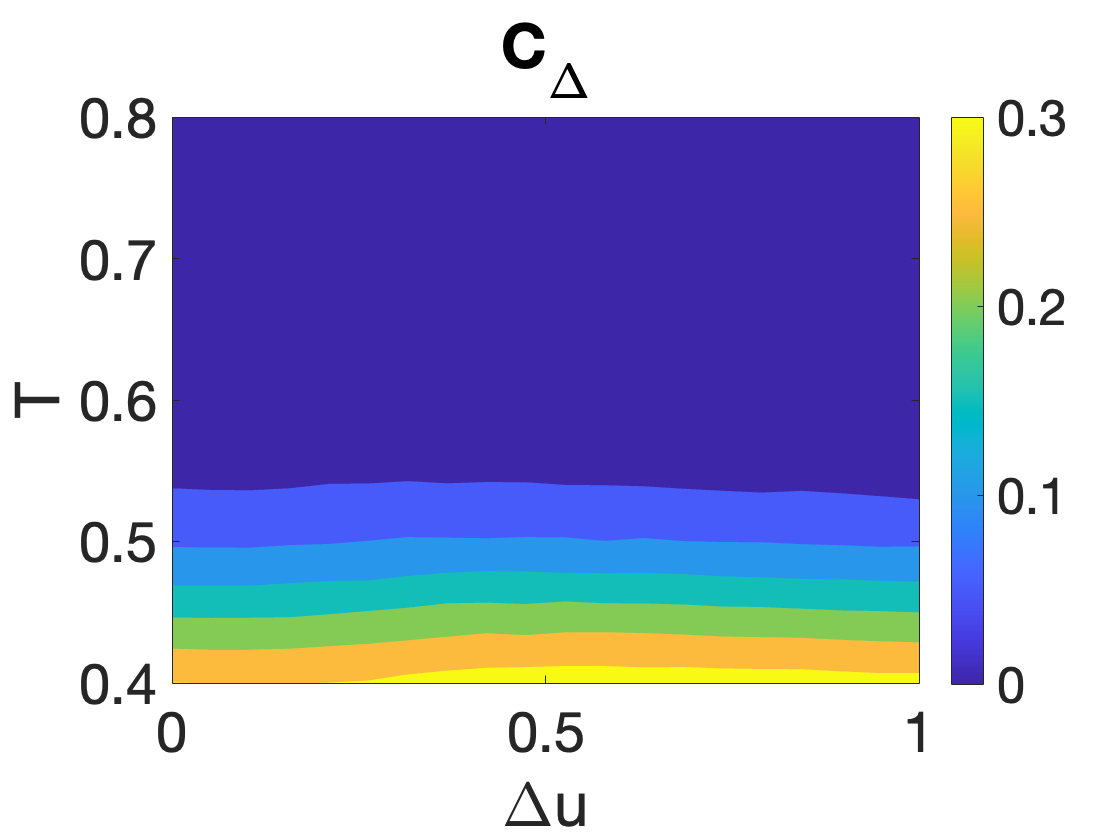}
		\includegraphics[width=4cm]{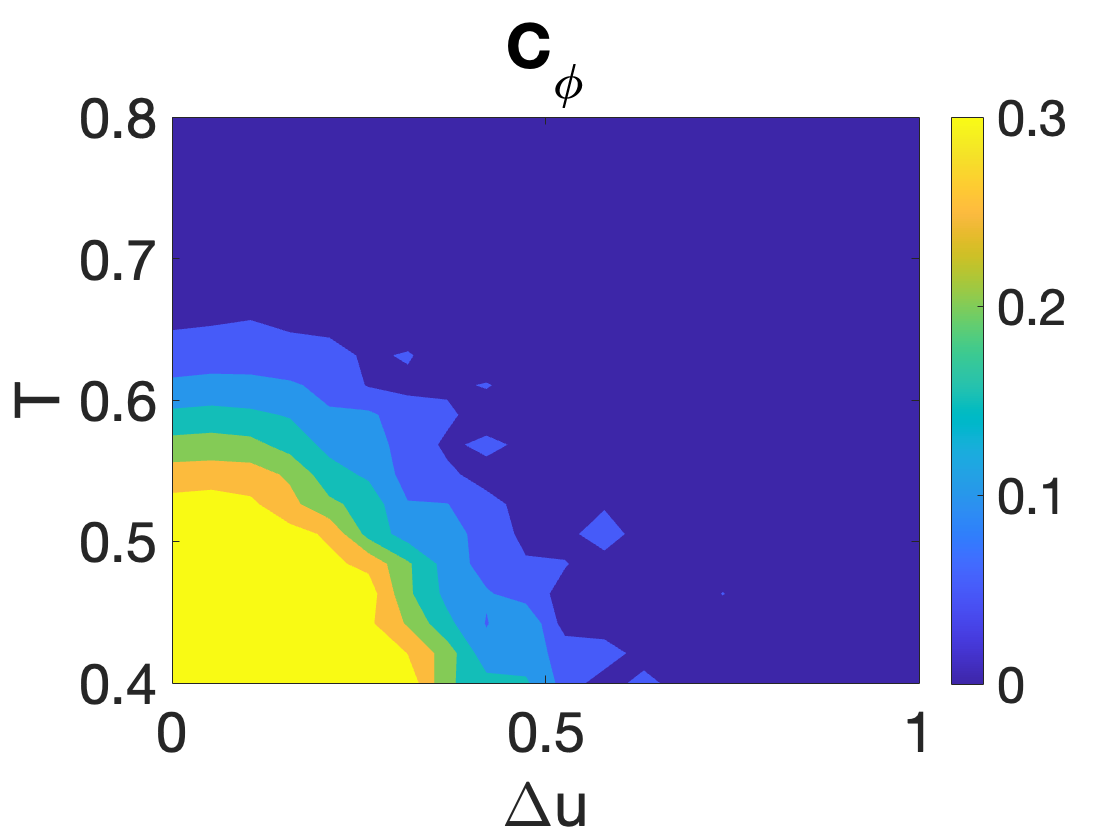}
		\caption{Correlation functions at the longest distance, for (left) charge-4e SC and (right) CDW, as a function of disorder strength $D$ and temperature $T$.}
		\label{F5} 
	\end{figure}
	
	Let us start with zero disorder $D=0$. At zero temperature (without thermal fluctuation), SC should be a stronger order as we choose $|u_\Delta|>|u_\phi|$. With thermal fluctuation (Fig.\ref{F5}), the commensurate CDW has a higher $T_c$, since the discrete symmetry breaking suffers from less fluctuation. This weak disorder limit corresponds to the phase diagram in the middle panel of Fig.\ref{F4}, which has a vestigial CDW phase. 
	
	As a strong disorder suppresses the CDW, the charge-4e SC has a higher $T_c$. This strong disorder limit corresponds to the left panel of Fig.\ref{F4}, with a vestigial charge-4e phase. There is a tetracritical point (crossing point between the two continuous phase transitions) in the disorder-temperature phase diagram, separating the above two limits. 
	In sum, CDW disorder can be crucial for stabilizing the vestigial charge-4e phase in a general commensurate system.

	\section{Discussion}
	In this study, our focus is specifically on the low-temperature surface properties of CsV$_3$Sb$_5$. Our analysis is solely based on the existing scanning tunneling microscopy results without making any additional assumptions on the 3D wave vectors of the CDW and PDW. On the surface of CVS, the PDW state is found to be non-degenerate, implying that there is only one unique approach to constructing a vestigial charge-4e phase. By utilizing the experimental STM data, we aim to provide a comprehensive understanding of the surface behavior and the specific characteristics of the vestigial charge-4e phase in CVS.
	
	A natural continuation of this study is to investigate the behavior of CsV$_3$Sb$_5$ in bulk. It is crucial to explore the existence of a three-dimensional PDW state and determine its out-of-plane wavevector to understand the symmetry-breaking phenomena. For instance, one may question whether the points $\pm \mathbf{Q}_i$ remain equivalent in the three-dimensional case.
	
	Observations have shown that the $4a_0$-CDW is confined to the Sb surface \cite{zhao2021,chen2021,liang2021}, indicating that it is unlikely to be a bulk effect. Consequently, the PDW states at wavevectors $\mathbf{Q}_{1}$ and $\mathbf{Q}_{2}$ are expected to differ. In the bulk, with the presence of more than two superconducting order parameters, multiple scenarios can give rise to vestigial charge-4e phases. By constructing the corresponding surface phase diagram based on a given bulk phase diagram, STM probes can play a crucial role in ruling out certain possibilities in the bulk. A specific example illustrating this process can be found in the appendix of this paper. However, it is important to note that our analysis of the bulk PDW wavevectors involves certain assumptions, as there is currently a lack of experimental results regarding the 3D wave vectors of the PDW in CsV$_3$Sb$_5$.
	
    Although this study primarily focuses on the Kagome superconductor CsV$_3$Sb$_5$, similar investigations could be extended to the other two Kagome superconductors, KV$_3$Sb$_5$ and RbV$_3$Sb$_5$. In KV$_3$Sb$_5$ and RbV$_3$Sb$_5$, the $2a_0$ CDW has been observed through various experimental techniques \cite{yin2021,ortiz2021Super,jiang2021,li2022,Li2021,Miao2021,liu2021,luo2022}. The unidirectional $4a_0$ CDW observed in CsV$_3$Sb$_5$ is not detected in these two materials. To date, no PDW has been observed in KV$_3$Sb$_5$ and RbV$_3$Sb$_5$, but there have been reports of dispersive scattering wave vectors at $\bf Q_i$ in the normal state of KV$_3$Sb$_5$ \cite{li2023}. 
    
    To fully understand the superconducting states in these two materials, it is thus crucial to check the existence of any non-dispersive CDW peaks at low temperature. For instance, if the dispersive scattering in KV$_3$Sb$_5$ became non-dispersive at $T_{CDW}$, this would suggest the existence of PDW in KVS. Interestingly, due to the presence of the $2a_0$ CDW in that system, CDW peaks at $\pm Q_1$ are equivalent. So the conventional uniform vestigial superconductivity $\Delta_{Q_1}\Delta_{-Q_1}$ should not exist. The correct form of the vestigial superconductivity thus depends on results from future experiments.

	{\it Acknowledgments - }We thank S. Raghu, S. Kivelson, D. Agterberg, and Z. Han for helpful discussions.
	
	\bibliography{citation}
	\appendix
	
	\section{Methods}
	In this section, we show the detailed steps for the classical Monte Carlo calculation.
	
	We start with a $L\times L$ square lattice with periodic boundary conditions. At each site $\bf r$, we initialize the CDW disorder $h({\bf r})$. The disorder is chosen independently from a uniform distribution $[-D, D]$. We also initialize a random configuration for $\Delta({\bf r})$ and $\phi({\bf r})$. Here, $\Delta({\bf r})$ are complex variables while $\phi({\bf r})$ are real variables.
	
	We can now compute the free energy, for a given $\Delta({\bf r})$ and $\phi({\bf r})$ configuration. The kinetic terms live on the bond. For example, the kinetic term for $\phi$ is $\sum_{\langle {\bf r r'}\rangle}\frac{k_\phi}{2}(\phi({\bf r})-\phi({\bf r'}))^2$, summed up nearest neighbored sites $\langle {\bf r r'}\rangle$. The other terms live on sites. For example, the disorder term is $\sum_{\bf r}h({\bf r})\phi({\bf r})$, summed up all sites.
	
	We now propose a random change for $\phi({\bf r})$ at a particular site ${\bf r}$. If the change in the free energy $\Delta E$ is negative, then this change is accepted. Otherwise, its acceptance rate is $\exp(-\Delta E/T)$. And then we move to another site. As we go through all $\phi({\bf r})$, we will perform the same consideration for the complex variables $\Delta({\bf r})$. As we go through all $\phi({\bf r})$ and $\Delta({\bf r})$, we finish one `step' in the simulation.
	
	After a sufficient number of steps, the system is stabilized and we are ready to perform measurements, as written in the main text. Two measurements need to be separated by a sufficient number of steps, to make them independent. By averaging out these measurements, we obtain the thermal average $\langle ... \rangle$.
	
	In the calculations, we take $L=20$, $k_{\Delta}=k_{\phi}=1$, $u_\Delta=-1$, $u_\phi=-0.95$ and $\lambda=0.2$. Classical Monte Carlo calculation is performed with $10^6$ steps and $10^3$ measurements. 
	
	We generate different 40 disorder configurations. By averaging their correlation function, we obtain the final disorder average. 
	
	\section{An example for the bulk}
	In the main text, we have identified three distinct surface phase diagrams that can be differentiated using STM probes. In this section, we outline the methodology for deriving a surface phase diagram based on a given bulk phase diagram.
	
	Since the 3D wave vectors of the bulk PDW are unknown, we have to make some assumptions about them. For simplicity, we assume six PDW $\pm\mathbf{Q}_{1,2,3}$ with zero out-of-plane wave vectors. Due to the lack of the $4a_0$ CDW in the bulk, $\mathbf{Q}_{1,2}$ are no longer equivalent points. And $\mathbf{Q}_{3}$ is unequivalent to $\bf 0$.
	
	\begin{figure}[h]
		\centering
		\includegraphics[width=8cm]{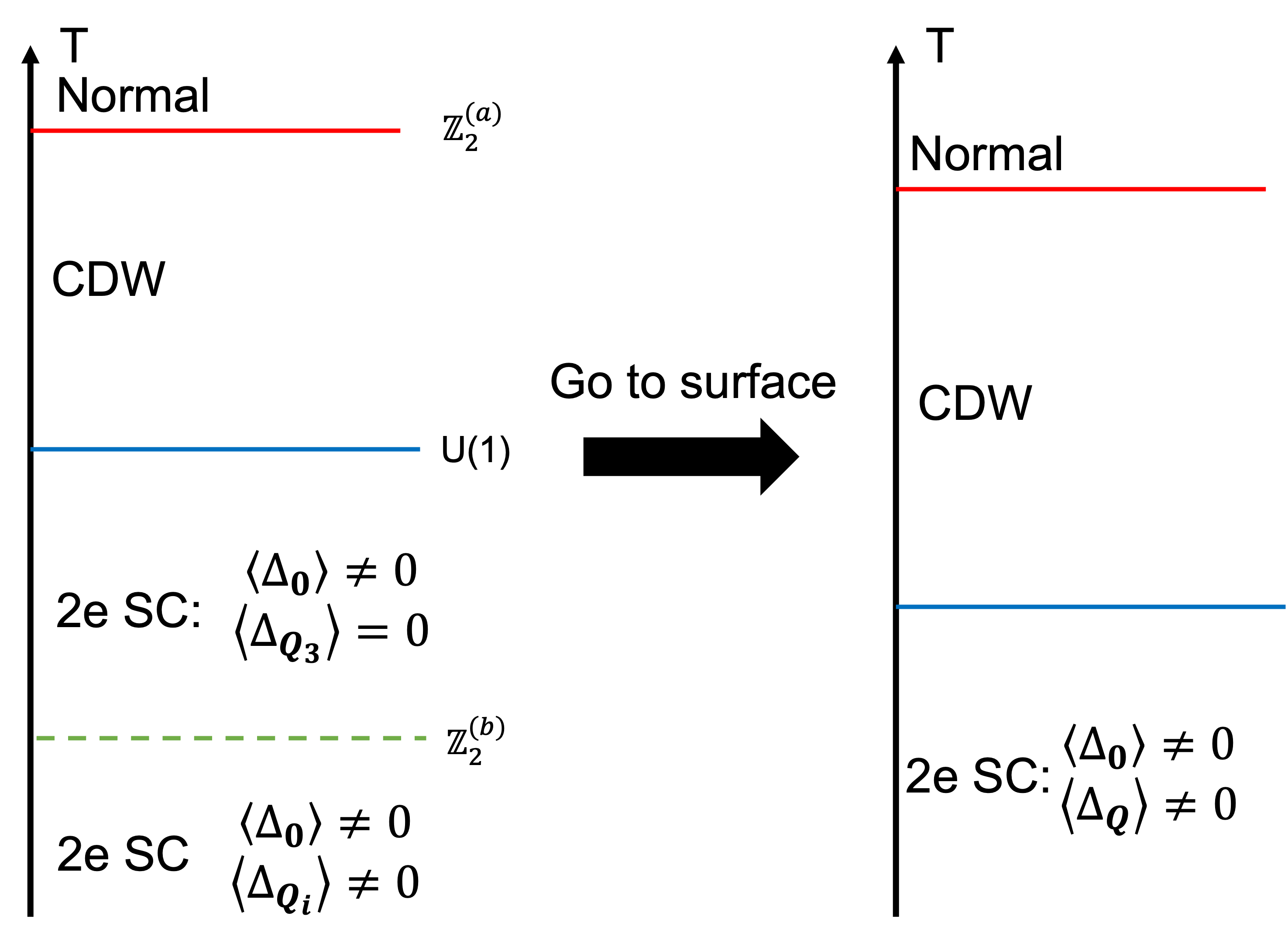}
		\caption{An example of obtaining the surface phase diagram from the bulk phase diagram.}
		\label{APF1} 
	\end{figure}
	Due to the $2a_0$ CDW (no matter whether it is $2\times2\times2$ or $2\times2\times4$), $\pm\mathbf{Q}_i$ are equivalent points because of the above assumptions. Compared with the surface, the bulk does not have the $4a_0$ CDW. So the total symmetry is larger. It becomes U(1)$\times\mathbb{Z}_2\times\mathbb{Z}_2$. Here U(1) describes the superconductivity. The first $\mathbb{Z}_2$ symmetry (denoted as $\mathbb{Z}_2^{(a)}$ below) is the same as in the main text. The second $\mathbb{Z}_2$ symmetry (denoted as $\mathbb{Z}_2^{(b)}$ below) corresponds to the $4a_0$ CDW on the surface.
	
	Now it is clear how to get a surface phase diagram from a bulk phase diagram. One only needs to further break an additional $\mathbb{Z}_2^{(b)}$ symmetry. 
	
	In the following example, the three symmetries are assumed to be broken at different temperatures. As we go to the surface phase diagram, the $\mathbb{Z}_2^{(b)}$ symmetry is no longer present due to the $4a_0$ CDW. The resulting surface phase diagram is the middle one in Fig.\ref{F4} in the main text.

	In the presence of multiple PDW states in the bulk, there exists a variety of possible bulk phase diagrams. However, by employing the STM probe technique proposed in the main text, it becomes feasible to identify the correct surface phase diagram. This, in turn, provides valuable insights into the potential bulk phase diagrams.

\end{document}